\documentclass[onecolumn,floatfix,superscriptaddress,showpacs,showkeys,nofootinbib]{revtex4}

\usepackage{epsfig}
\usepackage{amssymb,latexsym,amsmath}
\newcommand{\eq}[1]{\begin{align} #1 \end{align}}

\begin{document}
\title{Particle Number Fluctuations in the Microcanonical Ensemble}

\author{V.V. Begun}
\affiliation{Bogolyubov Institute for Theoretical Physics, Kiev, Ukraine}
\author{M.I. Gorenstein}
\affiliation{Bogolyubov Institute for Theoretical Physics, Kiev, Ukraine}
\affiliation{Institut f\"ur Theoretische Physik, Universit\"at Frankfurt,
Germany} \affiliation{Frankfurt Institute for Advanced Studies, Frankfurt,
Germany}
\author{A.P. Kostyuk}
\affiliation{Bogolyubov Institute for Theoretical Physics, Kiev, Ukraine}
\affiliation{Institut f\"ur Theoretische Physik, Universit\"at Frankfurt,
Germany} \affiliation{Frankfurt Institute for Advanced Studies, Frankfurt,
Germany}
\author{O.S. Zozulya}
\affiliation{Bogolyubov Institute for Theoretical Physics, Kiev, Ukraine}
\affiliation{Utrecht University, Utrecht, The Netherlands}

\begin{abstract}
Particle number fluctuations are studied in the microcanonical ensemble.
For the Boltzmann statistics we deduce exact analytical formulae for the
microcanonical partition functions in the case of non-interacting massless
neutral particles and charged particles with zero net charge. The particle
number fluctuations are calculated and we find that in the microcanonical
ensemble they are suppressed in comparison to the fluctuations in the
canonical and grand canonical ensembles. This remains valid in the
thermodynamic limit too, so that the well-known equivalence of all
statistical ensembles refers to average quantities, but does not apply to
fluctuations. In the thermodynamic limit we are able to calculate the
particle number fluctuations in the system of massive bosons and fermions
when the exact conservation laws of both the energy and charge are taken
into account.
\end{abstract}

\pacs{24.10.Pa, 05.20.-y}
\keywords{thermal model, microcanonical ensemble, fluctuations, thermodynamic
limit}

\maketitle

\section { Introduction}
 The statistical hadron gas
model (see e.g. Ref.~\cite{stat-model} and recent review \cite{PBM})
appears to be rather successful in describing the data of nucleus-nucleus
(A+A) collisions for particle multiplicities in a wide range of the
collision energies. Usually one considers a thermal system created in A+A
collision in the grand canonical ensemble (GCE). There are, however,
situations when the canonical ensemble (CE) \cite{ce} or even
microcanonical ensemble (MCE) \cite{mce} with explicit treatment of charge
conservations or both charge and energy conservations are required.
This happens, for example, when the statistical model is applied to elementary
$pp,~ p\overline{p},~ e^{+}e^{-}$ collisions. Different statistical ensembles are not
equivalent for small systems created in these collisions.

In A+A collisions one prefers to use the GCE because it is the most
convenient one from the technical point of view and due to the fact that
both the CE and MCE are equivalent to the GCE in the thermodynamic limit
(i.e. when the size of the system tends to infinity). However, the
thermodynamic equivalence of ensembles means only that the average values
of different physical quantities calculated in different ensembles are
equal to each other in the thermodynamic limit. On the other hand, the
analysis of fluctuations is also an important tool to study a physical
system. Event-by-event analysis of A+A collisions (see e.g.
Ref.~\cite{fluc}) can reveal new physical effects not seen in observables
averaged over a large sample of events.
 An essential part of the total fluctuations measured on the
event-by-event basis is expected to be the thermal fluctuations. It was
demonstrated for the first time
 in Ref.~\cite{ce-fluc} that particle number
fluctuations are different in the CE and GCE even in the thermodynamic
limit. In the present paper we extend the results of Ref.~\cite{ce-fluc}
and make the analytical calculations of the particle number fluctuations
in the MCE.
In the textbooks of statistical mechanics the particle number fluctuations
in the CE and MCE were not considered since the discussion was usually
limited to the non-relativistic cases and the number of particles are
assumed to be fixed in these ensembles. In the relativistic situation one
can only fix the conserved charges (in the CE), or both the energy and
conserved charges (in the MCE), while the particle numbers still fluctuate
both in the CE and MCE. Results of both Ref.~\cite{ce-fluc} and the
present paper demonstrate that the particle number fluctuations are
different in various ensembles even in the thermodynamic limit.

 The paper is organized in the following way. First, we deduce and study the
 exact analytical expressions in the MCE for massless neutral particles with
 Boltzmann statistics: the partition function and average number of
 particles in Sec.~II and particle number fluctuations in Sec.~III.
 The extension of these results for the system of
 charged particles with zero net charge  is considered in Sec.~IV.
 In Sec.~V we use the method of microscopic correlator proposed
 in Ref.~\cite{steph}.  This gives us a possibility to study
 the MCE fluctuations in the thermodynamic limit
 for a much more general situation. It includes the effects of quantum
 statistics and non-zero particle mass.  We consider the system of
 neutral particles and then extend the formulation to
 charged particles too. Both the energy and charge exact conservation laws
are imposed. We summarize our consideration
 in Sec.~VI.

 \section  {The Microcanonical partition function and average number of particles}
In order to calculate analytically the microcanonical partition
function\footnote{We define the microcanonical ensemble as the statistical
system with fixed energy. An exact conservation of momentum, angular
momentum and parity are neglected here. The MCE with conserving momentum
was considered in Ref.~\cite{mce}.} we start in Secs.~II-III with the
system of non-interacting massless neutral particles and neglect the
effects of quantum statistics (the extensions will be treated in
Secs.~IV-V).
The microcanonical partition function for one massless
particle with energy $E$ in the volume $V$ can be easily
calculated:
\begin{align}\label{omega1}
 W_1(E,V)
 \;=\; \frac{g\;V}{(2\pi)^3} \int d^3p~\delta(E-|\vec{p}|)
 \;=\; 4\pi\, \frac{g\;V}{(2\pi )^3}
 \int_0^{\infty} dp\;p^2\;\delta(E-p)
 \;=\; \frac{g\;V}{2\pi^2}\; E^2\;.
\end{align}
Here $\;g\;$ is the degeneracy factor. In the case of two
massless particles:
\begin{align}\label{omega2}
W_2(E,V)
 &\;=\; \frac{1}{2}\,\frac{gV}{(2\pi )^3}
 \int d^3q~ \frac{gV}{(2\pi )^3}
       \int d^3p\;\delta(E-|\vec{q}|-|\vec{p}|) \nonumber
 \\
 &\;=\; \frac{1}{2}\,4\pi\,\frac{gV}{(2\pi )^3}
       \int_0^E dq\,q^2\;W_1(E-q,V)
       \nonumber
  \\
 &\;=\; \frac{1}{2} \left(\frac{gV}{2\pi^2}\right)^2
        \int_0^E dq\,q^2\;(E-q)^2 \;=\; \frac{1}{60}
        \left(\frac{gV}{2\pi^2}\right)^2 E^5~.
\end{align}
The factor $1/2$ appears because particles are identical.
It can be proven that the $N$-particle microcanonical
partition function $W_N(E,V)$ has
the form (see Appendix A):
\begin{align}\label{omegaNa}
W_N(E,V)
 \;=\;
 \frac{1}{E}\,\frac{x^N}{(3N-1)!N!}~,
\end{align}
where
\begin{align}\label{xdef}
x \equiv gVE^3/\pi^2~.
\end{align}

The total partition function in the MCE is:
\begin{align}\label{WE}
W(E,V) &\;\equiv \; \sum_{N=1}^{\infty}W_N(E,V)
      \;=\; \frac{1}{E}\sum_{N=1}^{\infty}\frac{x^N}{(3N-1)!N!}
      \;=\;
      \frac{x}{E}\sum_{n=0}^{\infty}\frac{x^n}{(3n+2)!(n+1)!}
      \\
     &\;=\; \frac{x}{E}\sum_{n=0}^{\infty} \frac{3x^n}{(3n+3)! n!}
      \;=\; \frac{x}{2E}\sum_{n=0}^{\infty}
       \frac{(x/27)^n}{\left(\frac{4}{3}\right)_n
            \left(\frac{5}{3}\right)_n (2)_n~n!}
            \nonumber
      \\
     &\;=\; \frac{x}{2E}\;\;_0F_3
     \left(;\,\frac{4}{3},\frac{5}{3},2;\,\frac{x}{27}\right)~,\nonumber
\end{align}
where we
%
%
use (see e.g. Ref.~\cite{I} and Appendix B) the Pochhammer symbol $(a)_n$ (\ref{Poch})
and the generalized hypergeometric function
(GHF) $_pF_q(a_1,...,a_p;b_1,...,b_q;z)$ (\ref{GHF}).

%
In the GCE the independent variables are the volume $V$ and temperature
$T$. For neutral massless particles with Boltzmann statistics one easily
finds the average number of particles and average energy:
\begin{align}
\langle N\rangle_{g.c.e.}
 \;=\; \frac{gVT^3}{\pi^2}\;,~~~~
 \langle E\rangle_{g.c.e.}
 \;=\;  \frac{3gVT^4}{\pi^2}\;. \label{Egce}
\end{align}
We want to compare the results of the MCE and GCE at equal volumes $V$ and
energies $\langle E\rangle_{g.c.e.} = E$. From Eq.~(\ref{Egce}) it
follows:
\begin{align}\label{Ngce1}
\langle N\rangle_{g.c.e.} ~\equiv~\overline{N}\;=\;
\left(\frac{x}{27}\right)^{1/4}\;.
\end{align}
The average number of particles in the MCE equals to:
\begin{align}\label{Nmce}
\langle N\rangle_{m.c.e.}
 &\;\equiv\; \frac{1}{W(E,V)}\sum_{N=1}^{\infty} N\,W_N(E,V)
  \;=\; \frac{1}{W(E,V)}\frac{1}{E}\sum_{N=1}^{\infty}
        \frac{x^N}{(3N-1)!(N-1)!}
  \\
 &\;=\; \frac{1}{W(E,V)}\frac{x}{E}\sum_{n=0}^{\infty}
        \frac{x^n}{(3n+2)!n!}\;
 =\; \frac{1}{W(E,V)}\frac{x}{2E}
        \sum_{n=0}^{\infty} \frac{(x/27)^n}
        {(1)_n\left(\frac{4}{3}\right)_n\left(\frac{5}{3}\right)_n n!}\nonumber\\
  &\;=\; \frac{1}{W(E,V)}\,\frac{x}{2E}\;\;_0F_3
        \left(;\,1\,,\;\frac{4}{3}\,,\;\frac{5}{3}\,;\;\frac{x}{27}\right)
  \;=\; \frac{_0F_3
        \left(;\,1,\,\frac{4}{3},\,\frac{5}{3};\,\frac{x}{27}\right)}
        {_0F_3
        \left(;\,\frac{4}{3},\,\frac{5}{3},\,2;\,\frac{x}{27}\right)}\;.\nonumber
\end{align}

\begin{figure}[t]
\begin{center}
\epsfig{file=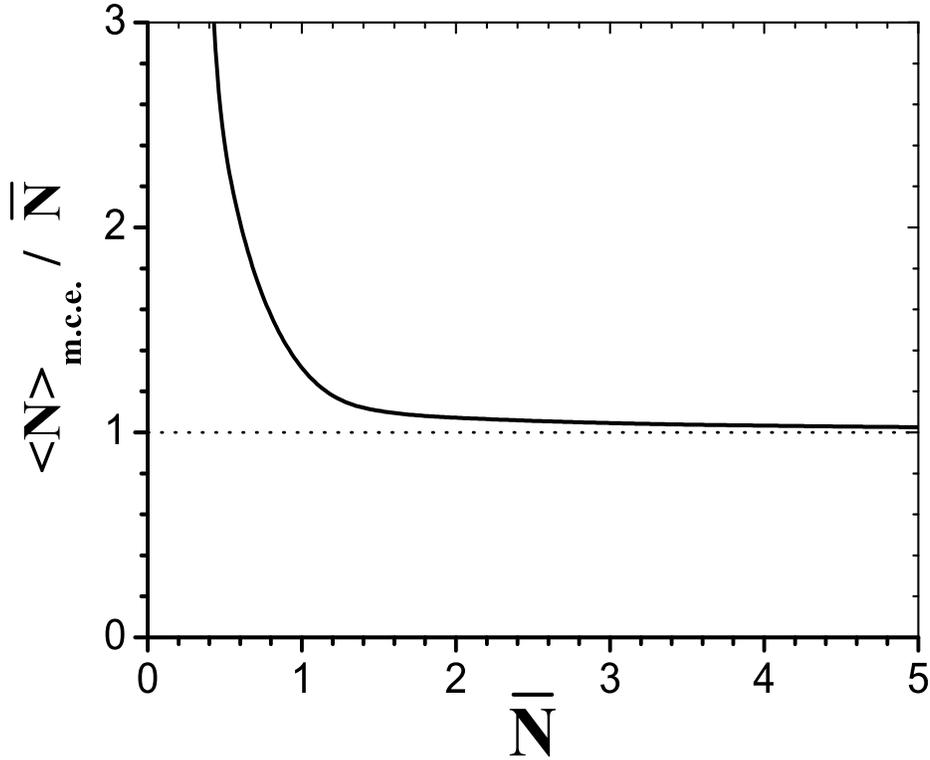,height=12cm,width=14cm} \vspace{-1cm}
 \caption{The ratio of average particle number
$\langle N\rangle_{m.c.e.}$ (\ref{Nmce}) in the MCE to that $\overline{N}$
(\ref{Ngce1}) in the GCE. \label{fig1}} \vspace{0.5cm}
\end{center}
\end{figure}

From Eq.~(\ref{Nmce}), using the asymptotic behavior of the GHF
(\ref{GHFas}), one finds the asymptotic expansion of $\langle
N\rangle_{m.c.e.}$ (\ref{Nmce}) at $\overline{N}\rightarrow\infty$:
\begin{align}\label{Nmce2}
\langle N\rangle_{m.c.e.}~\simeq~\overline{N}\left(1~+
~\frac{1}{8~\overline{N}}~+~\frac{35}{1152~\overline{N}^2}~+~
...\right)~.
\end{align}
The MCE and GCE are thermodynamically equivalent, i.e. $\langle
N\rangle_{m.c.e.}=\overline{N}$, in the thermodynamic limit $VE^3
\rightarrow \infty$. The dependence of the ratio $\langle
N\rangle_{m.c.e.}/\overline{N}$ on $\overline{N}$ is shown in
Fig.~\ref{fig1}.

As seen from Fig.~\ref{fig1} the ratio $\langle
N\rangle_{m.c.e.}/\overline{N}$ goes to 1 at $\overline{N}\gg 1$. As the
smallest number of particles in the MCE equals to 1, one finds that
$\langle N\rangle_{m.c.e.}\simeq 1$ when the volume $V$ and the energy $E$
of the system become {\it small}, i.e. if $\;VE^3\ll 1\;$. On the other
hand, in the GCE one finds $\langle N\rangle_{g.c.e.}\propto
(VE^3)^{\frac{1}{4}}\ll 1$ in this limit of small systems. Therefore, at
$\overline{N}<1$ the ratio $\langle N\rangle_{m.c.e.}/\overline{N}$ is
larger than 1 and it increases monotonously when $\overline{N}\rightarrow
0$.
\section{Particle number fluctuations}
\begin{figure}[b]
\begin{center}
\epsfig{file=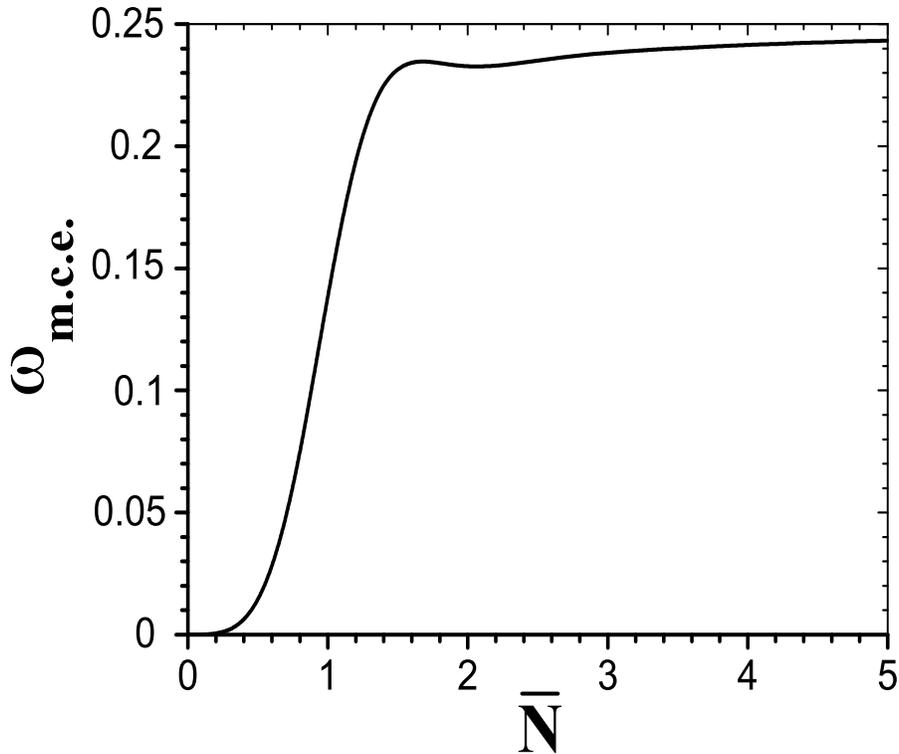,height=12cm,width=14cm} \vspace{-1cm}
 \caption{The scaled variance $\omega_{m.c.e.}$
 (\ref{omegamce}) in the MCE. \label{fig2}}
\vspace{0.5cm}
\end{center}
\end{figure}

To study the particle number fluctuations in the MCE we calculate
\begin{align}\label{N^2mce}
\langle N^2\rangle_{m.c.e.}
 &\;\equiv\; \frac{1}{W(E,V)}\sum_{N=1}^{\infty} N^2\,W_N(E,V)
  \;=\; \frac{1}{W(E,V)}\frac{1}{E}\sum_{N=1}^{\infty}
        \frac{N^2x^N}{(3N-1)!N!}
  \\
 &\;=\; \frac{1}{W(E,V)}\frac{1}{E}\;\left[\; \frac{x}{2}\;\;
        _0F_3\left(;\,1,\frac{4}{3},\frac{5}{3};\,\frac{x}{27}\right)
  \;+\; \frac{x^2}{120}\;\;
        _0F_3\left(;\,2,\frac{7}{3},\frac{8}{3};\,\frac{x}{27}\right)
        \;\right]\nonumber
  \\
 &\;=\; \langle N\rangle_{m.c.e.}
  \;+\; \frac{x}{60}\;\frac{
  _0F_3\left(;\,2,\frac{7}{3},\frac{8}{3};\,\frac{x}{27}\right)}{
  _0F_3\left(;\,\frac{4}{3},\,\frac{5}{3},\,2;\,\frac{x}{27}\right)}\;.\nonumber
\end{align}
The asymptotic expansion of $\langle N^2\rangle_{m.c.e.}$ (\ref{N^2mce})
can be found using Eq.~(\ref{GHFas}):
\begin{equation}\label{n2as}
\langle N^2\rangle_{m.c.e.} ~\simeq ~\overline{N}^2\left(1~+
~\frac{1}{2~\overline{N}}~+~\frac{11}{144~\overline{N}^2}+~ \dots
\right)~.
\end{equation}
A measure of the fluctuations, the scaled variance $\omega$, is defined as
usual,
\begin{align} \label{omegamce}
\omega_{m.c.e.}
 \;=\; \frac{\langle N^2\rangle_{m.c.e.} - \langle N\rangle_{m.c.e.}^2}
       {\langle N\rangle_{m.c.e.}}\;,
\end{align}
and it is plotted in Fig.~\ref{fig2}.

\begin{figure}[p]
\epsfig{file=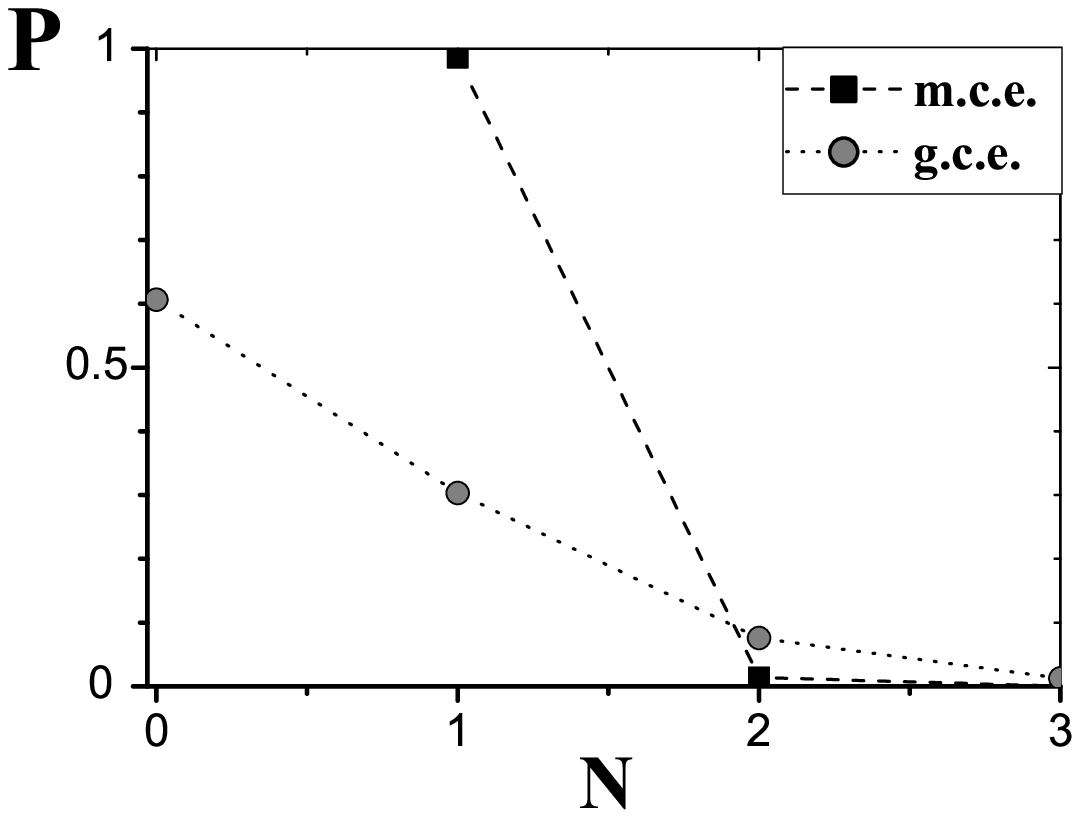,height=12cm,width=14cm}
 \vspace{-1.1cm}
 \caption{The particle number distributions in the MCE
 (\ref{PNmce}) and  GCE (\ref{PNgce})
 for $~\overline{N}=0.5~$. \label{fig3}}
 \hspace{-0.67cm}
\epsfig{file=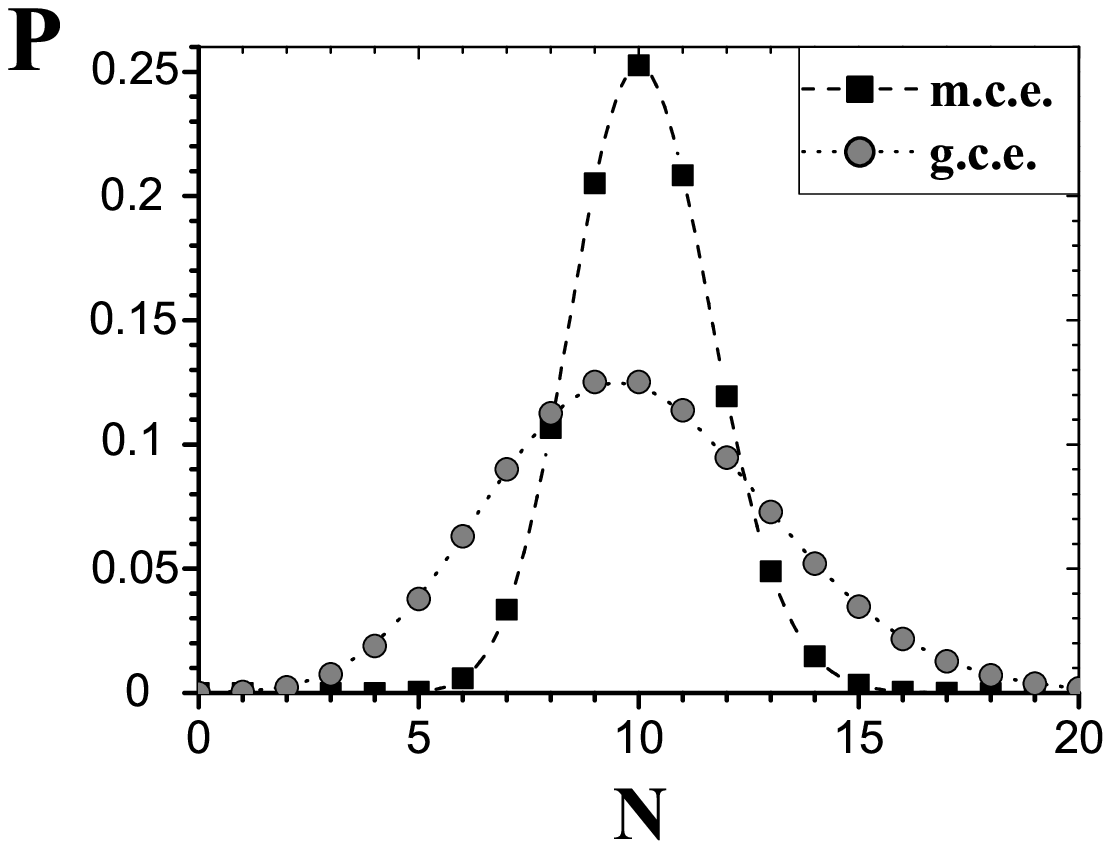,height=12cm,width=14.9cm}
 \vspace{-1.1cm}
 \caption{The same as in Fig.~\ref{fig3}, but
 for $~\overline{N}=10~$. \label{fig4}}
\end{figure}

The asymptotic expansion of $\omega_{m.c.e.}$
(\ref{omegamce}) has the form:
\begin{equation}\label{omegas}
\omega_{m.c.e.} \simeq \frac{1}{4}
\left(1~-
~\frac{1}{8~\overline{N}}~+~   \dots \right)~.
\end{equation}

In Figs.~\ref{fig3} and \ref{fig4} we present the particle number
distributions in the MCE and GCE for small $(\overline{N}=0.5)$ and large
$(\overline{N}=10)$ systems.

 In the MCE the particle number distribution equals
\begin{align}\label{PNmce}
P_{m.c.e.}(E,V,N)~\equiv~\frac{W_{N}(E,V)}{W(E,V)} \;=\; \frac{1}{E\,
W(E,V)}\;\frac{x^N}{(3N-1)!N!}\;,\qquad N=1,\;2,\ldots
\end{align}
and it has the Poisson form in the GCE
\begin{align}\label{PNgce}
P_{g.c.e.}(\overline{N},N) \;=\;
 \exp\left(-~\overline{N}\right)~\frac{\overline{N}^{N}}{N!}\;,
  \qquad N=0,\;1,\;2,\ldots
\end{align}
 Note
that Poisson distribution $P_{g.c.e.}(\overline{N},N)$ (\ref{PNgce})
results in
\begin{align}\label{omegagce}
\omega_{g.c.e.}~\equiv
~\frac{\overline{N^{2}}~-~\overline{N}^{2}}{\overline{N}}~=~1~.
\end{align}
It is seen from Eq.~(\ref{omegas}) that in the thermodynamic limit $V
\rightarrow \infty$ the MCE scaled variance equals to
$\omega_{m.c.e.}=1/4$, and it remains quarter as large as the scaled
variance $\omega_{g.c.e.}$ (\ref{omegagce}) in the GCE. We conclude that
the particle number distributions are different in the MCE and GCE. The
MCE particle number distribution (\ref{PNmce}) is narrower than that of
the GCE (\ref{PNgce}) in both small ($\overline{N}<1$) and large
($\overline{N}>1$) systems. At $\overline{N}<1$ the probability
distributions $P_{m.c.e.}(E,V,N)$ (\ref{PNmce}) and
$P_{g.c.e.}(\overline{N},N)$ (\ref{PNmce}) both have their maximums at the
smallest values of $N$. Then the crucial difference between two ensembles
follows from the fact that the minimal allowed value of $N$ is $N=0$ in
the GCE, but $N=1$ in the MCE (see Fig.~\ref{fig3}).

In the thermodynamic limit the average number of particles goes to
infinity and the main contribution to the microcanonical partition
function $W(E,V)$ (\ref{WE}) comes from the states with large number of
particles $N\gg 1$. In this limit the MCE particle number distribution can
be simplified in the following way. Using the Stirling formula for
factorials one finds:
\begin{align}\label{int1}
W_N(E,V)~=~\frac{1}{E}~\frac{x^N}{(3N-1)!~N!}
~\simeq~\frac{1}{E}~\exp[f(N)]~,
\end{align}
where
$
f(N)~\simeq~ N \log \left(\frac{x}{27}\right) - 4N(\log N - 1)
+\frac{1}{2} \log 3 - \log (2\pi)~.
$
We expand the function $f(N)$ in Taylor series near the point of its maximum
$\overline{N}$.
For $\overline{N} \gg 1$ and $|N-\overline{N}| \ll \overline{N}$ we find:
\begin{align}\label{int2}
f(N) ~\simeq~ f(\overline{N})
     +f^{\prime\prime}(\overline{N})~
     \frac{1}{2}~\left(N-\overline{N}\right)^2~
     =~ f(\overline{N})~-~
     \frac{2~\left(N-\overline{N}\right)^2}{\overline{N}}~,
\end{align}
where $\overline{N}=(x/27)^{1/4}$ is fixed by the condition
$f^{\prime}(\overline{N})=0$, and it coincides with the result of
Eq.~(\ref{Ngce1}). From Eq.~(\ref{int2}) it follows that the particle
number distribution in the MCE can be approximated as:
\begin{align}\label{gaussN}
P_{m.c.e.}(E,V,N) \;\propto \; \exp\left[~-~\frac{2\left(N -
\overline{N}\right)^2}
    {\overline{N}}\right]~.
\end{align}
For the Gauss distribution $P_{G}(N)\propto
\exp\left[-(N-\overline{N})^{2}/2\sigma^{2}\right]$ the variance is easily
calculated at $\overline{N}\rightarrow \infty$ and it equals to $\langle
N^{2}\rangle -\langle N\rangle^{2}=\sigma^2$. Therefore, from
Eq.~(\ref{gaussN}) it follows $\langle N^{2}\rangle_{m.c.e.} -\langle
N\rangle_{m.c.e.}^{2}=\overline{N}/4$, and the MCE scaled variance is
$\omega_{m.c.e.} = 1/4$.
The Poisson distribution (\ref{PNgce}) for $\overline{N}\gg 1$ can be also
approximated by the Gauss distribution, but it equals to
 $P_{g.c.e.}(\overline{N},N)\propto
\exp\left[-(N-\overline{N})^{2}/2\overline{N}\right]$, and this leads to
$\omega_{g.c.e.}=1$. Therefore, at $\overline{N}\gg1$ both the MCE and GCE
particle number distributions can be approximated by the Gauss
distributions with the same average value $\overline{N}$, but with
different widths: $\sigma^{2}_{m.c.e.}=\overline{N}/4$ and
$\sigma^{2}_{g.c.e.}=\overline{N}$. A consequence of this is that the MCE
 scaled variance is a quarter the size of that in the
GCE for classical massless neutral particles in the thermodynamic limit.
%
%
\section {The MCE for massless charged particles }
\begin{figure}[p]
 \epsfig{file=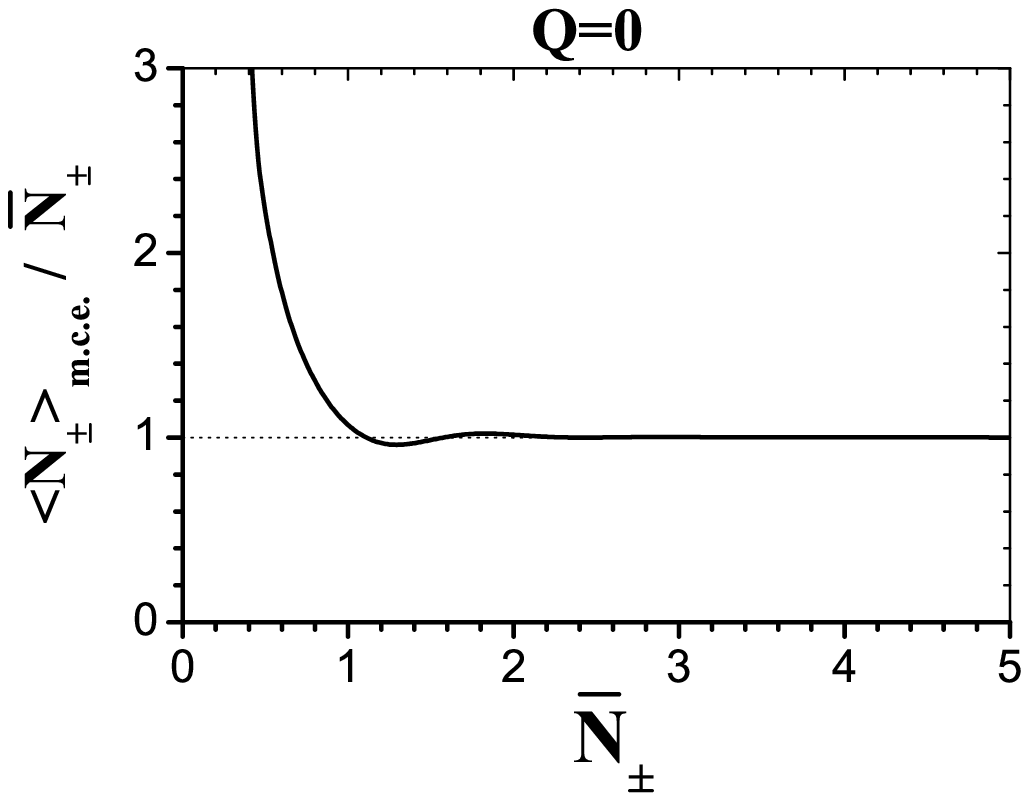,height=11.5cm,width=14cm}
 \vspace{-1cm}
 \caption{The ratio of average particle number
$\langle N_{\pm}\rangle_{m.c.e.}$ (\ref{Npmmce}) in the MCE to
that $\overline{N}_{\pm}$ (\ref{Npmgce}) in the GCE for positively
(negatively) charged
 particles at zero net charge $\;Q=0\;$. \label{fig5}}
\hspace{-1.2cm}
\epsfig{file=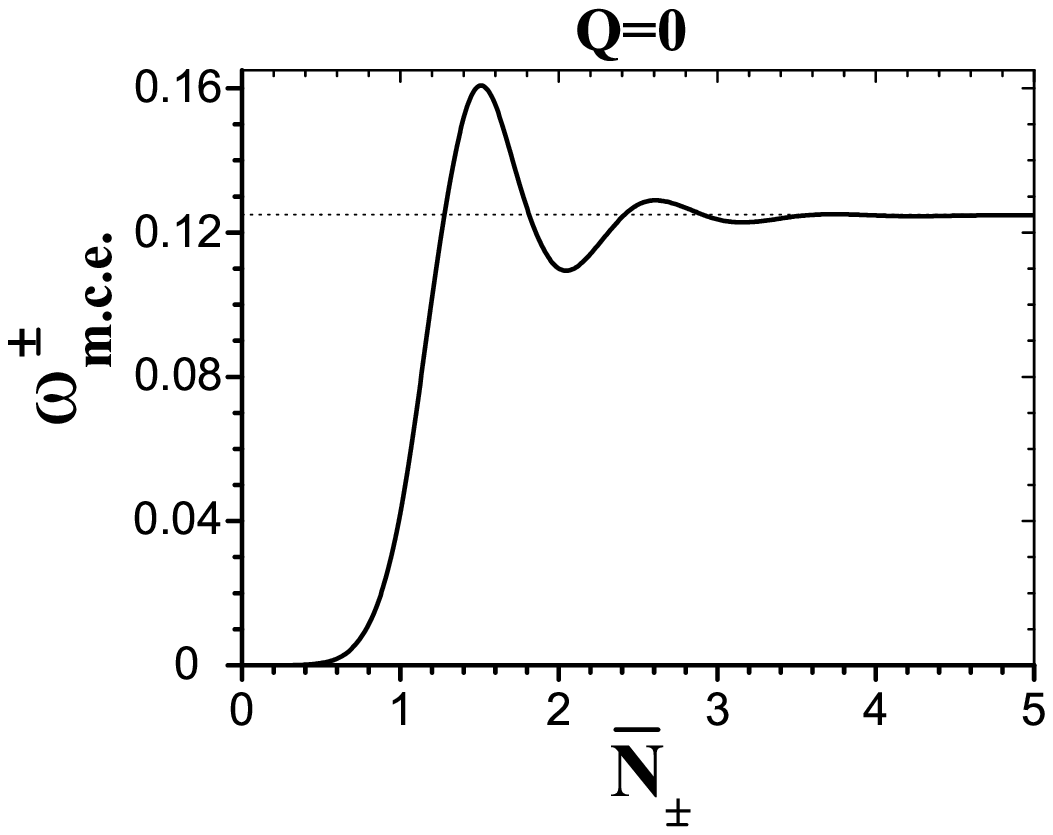,height=11.5cm,width=15.2cm}
 \vspace{-1cm}
 \caption{The scaled variance
$\omega_{m.c.e.}^{\pm}$ (\ref{omegamceQ}) for positively (negatively)
charged
 particles in the MCE at zero net charge $\;Q=0\;$.
\label{fig6} }
\end{figure}

The microcanonical partition function discussed in the previous sections
can be generalized for the system of charged particles. If the system net
charge $Q$ equals zero the number of positively charged $N_{+}$ and
negatively charged $N_-$ particles are equal in each microscopic
configuration. The total MCE partition function is ($a\equiv gV/\pi^2$):
\begin{align}\label{Wmce}
&W(E,V,Q=0) \;=\;
  \sum_{N_+=1}^{\infty}\sum_{N_-=1}^{\infty}
   \int_0^{\infty}dE_+ \int_0^{\infty} dE_-~
        W_{N_+}(E_+,V)\;W_{N_-}(E_-,V)\nonumber \\
  &\times      \delta(N_+-N_-)\;\delta\left[E-(E_++E_-)\right]
  \;=\;  \sum_{N_+=1}^{\infty}\int_0^{\infty}dE_+\;
         W_{N_+}(E_+,V)\;W_{N_+}(E-E_+,V) \nonumber \\
 & \;=\;
 \sum_{N_+=1}^{\infty} \frac{a^{2N_+}}{(3N_+-1)!^2\;N_+!^2}\;
 \int_0^E dE_+\; E_+^{3N_+-1}(E-E_+)^{3N_+-1}\;.
\end{align}
The last integral in Eq.~(\ref{Wmce}) can be easily evaluated
%
%
using the Euler Beta-function:
\begin{align}
 \int_0^E dE_+\; E_+^{3N_+-1}(E-E_+)^{3N_+-1}~=~
 E^{6 N_+ - 1} ~B(3 N_+,3 N_+) = E^{6 N_+ - 1} ~
 \frac{(3 N_+ - 1)!^2}{(6 N_+ - 1)!}.
\end{align}
Finally one finds:
%
%
\begin{align}\label{Wmce1}
& W(E,V,Q=0)
   \;=\; \sum_{N_+=1}^{\infty} \frac{a^{2N_+}\,E^{6N_+-1}}{(6N_+-1)!\,N_+!^2}
   \;=\; \frac{1}{E} \sum_{N_+=1}^{\infty}
   \frac{x^{2N_+}}{(6N_+-1)!\,N_+!^2} \nonumber
    \\
 & \;=\; \frac{x^2}{E} \sum_{n=0}^{\infty}\frac{x^{2n}}{(6n+5)!\,(n+1)!^2}
   \;=\; \frac{x^2}{120\,E}\;\;
       _0F_7\left(;\;\frac{7}{6},\,\frac{4}{3},\,\frac{3}{2},\,\frac{5}{3},
       \,\frac{11}{6},\,2,\,2;\,\left(\frac{x}{216}\right)^2\right)\;.
\end{align}
%
%
%
Similarly to Eqs.~(\ref{Nmce}, \ref{N^2mce}) after some calculations
one  obtains:
\begin{align}\label{Npmmce}
\langle N_{\pm}\rangle_{m.c.e.} \;=\; \frac{
 _0F_7\left(;\;1,\,\frac{7}{6},\,\frac{4}{3},\,\frac{3}{2},\,\frac{5}{3},
       \,\frac{11}{6},\,2;\,\left(\frac{x}{216}\right)^2\right)
       }{
 _0F_7\left(;\,\frac{7}{6},\,\frac{4}{3},\,\frac{3}{2},\,\frac{5}{3},
       \,\frac{11}{6},\,2,\,2;\,\left(\frac{x}{216}\right)^2\right)}\;,
\end{align}
and
\begin{align}\label{N2pmmce}
\langle N_{\pm}^2\rangle_{m.c.e.} \;=\; \frac{
 _0F_7\left(;\;1,\,1,\,\frac{7}{6},\,\frac{4}{3},\,\frac{3}{2},\,\frac{5}{3},
       \,\frac{11}{6};\,\left(\frac{x}{216}\right)^2\right)
       }{
 _0F_7\left(;\,\frac{7}{6},\,\frac{4}{3},\,\frac{3}{2},\,\frac{5}{3},
       \,\frac{11}{6},\,2,\,2;\,\left(\frac{x}{216}\right)^2\right)}\;.
\end{align}
In the GCE of Boltzmann massless charged particles with $~Q=\langle
N_+\rangle_{g.c.e.} - \langle N_-\rangle_{g.c.e.} =0\;$ one finds:
\begin{align}\label{gcecharge}
\langle N_{\pm}\rangle_{g.c.e.}
 \;=\; \frac{gVT^3}{\pi^2}\;,~~~~
  \langle E\rangle_{g.c.e.}
 \;=\; \frac{6gVT^4}{\pi^2}\;.
\end{align}
The results of the MCE and GCE are again compared at equal volumes $\;V\;$
and energies $~\langle E\rangle_{g.c.e.} = E\;$.
From Eq.~(\ref{gcecharge}) it follows:
\begin{align}\label{Npmgce}
\langle N_{\pm}\rangle_{g.c.e.}
~\equiv~ \overline{N}_{\pm}\;=\;
\left(\frac{x}{216}\right)^{1/4}\;.
\end{align}
The asymptotic expansions for $\langle N_{\pm}\rangle_{m.c.e.}$ (\ref{Npmmce}) and
$\langle N^2_{\pm}\rangle_{m.c.e.}$ (\ref{N2pmmce}) at
$\overline{N}_{\pm}\rightarrow \infty$ are found using Eq.~(\ref{GHFas}):
\begin{align}\label{Npmas}
\langle N_{\pm}\rangle_{m.c.e.} \;\simeq\; \overline{N}_{\pm}\left(
1~+~\frac{49}{2304 \overline{N}_{\pm}^2}~+~
\frac{49}{9216 \overline{N}_{\pm}^3}~+~...\right)
\end{align}
and
\begin{align}\label{N2pmas}
\langle N_{\pm}^2\rangle_{m.c.e.} \;\simeq\; \overline{N}^2_{\pm}\left(
1~+~\frac{1}{8 \overline{N}_{\pm}}~+~
\frac{49}{1152 \overline{N}_{\pm}^2}~+~
\frac{49}{6144 \overline{N}_{\pm}^3}~+~...\right)~.
\end{align}
The behavior of the ratio $\langle
N_{\pm}\rangle_{m.c.e.}/\overline{N}_{\pm}$ is shown in
Fig. \ref{fig5}.

The behavior of the scaled variance,
\begin{align}\label{omegamceQ}
\omega_{m.c.e.}^{\pm} ~\equiv~\frac{\langle
N_{\pm}^{2}\rangle_{m.c.e.}~-~\langle N_{\pm}\rangle_{m.c.e.}^{2}}{\langle
N_{\pm}\rangle_{m.c.e.}}\;,
\end{align}
is shown in Fig.~\ref{fig6}.
%
The asymptotic expansion of $\omega_{m.c.e.}^{\pm}$ at
$\overline{N}_{\pm}\rightarrow \infty$ has the form:
\begin{align}\label{omegamceQas}
\omega_{m.c.e.}^{\pm} ~\simeq~\frac{1}{8}~
\left(1~-~\frac{49}{1152 \overline{N}_{\pm}^2}~+~...
\right)~,
\end{align}
while $~\omega_{g.c.e.}^{\pm}=1\;$, the same as for neutral particles, and
$~\omega_{c.e.}^{\pm}=1/2\;$ \cite{ce-fluc}. From Eq.~(\ref{omegamceQas})
and Fig.~\ref{fig6} one sees that $\omega_{m.c.e.}^{\pm}=1/8$ in the
thermodynamic limit, and this is by a factor of $1/8$ smaller than the
scaled variance in the GCE and by a factor of $1/4$ than in the CE.
Therefore, for the system of massless particles with Boltzmann statistics
the exact energy conservation leads to the MCE suppression of the particle
number scaled variance in the thermodynamic limit by a factor of $1/4$,
and the exact charge conservation makes an additional suppression by a
factor of $1/2$.

From Eq.~(\ref{Wmce1}) we find the positive (negative) particle number
distribution in the MCE at net charge $Q$ equal to zero:
\begin{align}\label{PmceQ}
P_{m.c.e.}(E,V,N_{\pm},Q=0)~=~\frac{1}{E ~W(E,V,Q=0)}~
      \frac{x^{2N_{\pm}}}{(6N_{\pm}-1)!\,N_{\pm}!^2} ~.
 \end{align}
\begin{figure}[t]
 \hspace{-1.5cm}
\epsfig{file=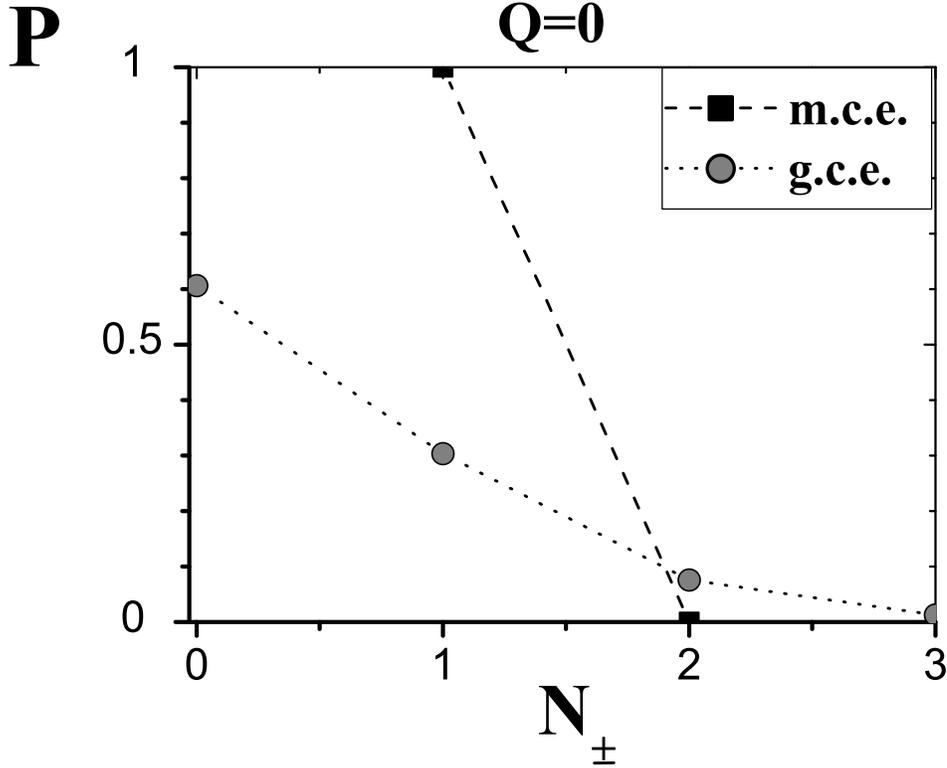,height=12cm,width=14cm}
 \vspace{-1.1cm}
 \caption{The particle number distributions
 of positively (negatively) charged particles in the MCE
 (\ref{PmceQ}) and in the GCE (the Poisson
 distribution) at zero net charge $~Q=~0$ and
 $\overline{N}_{\pm}=0.5$. \label{fig7}}
\end{figure}
\begin{figure}[t]
 \hspace{-0.5cm}
\epsfig{file=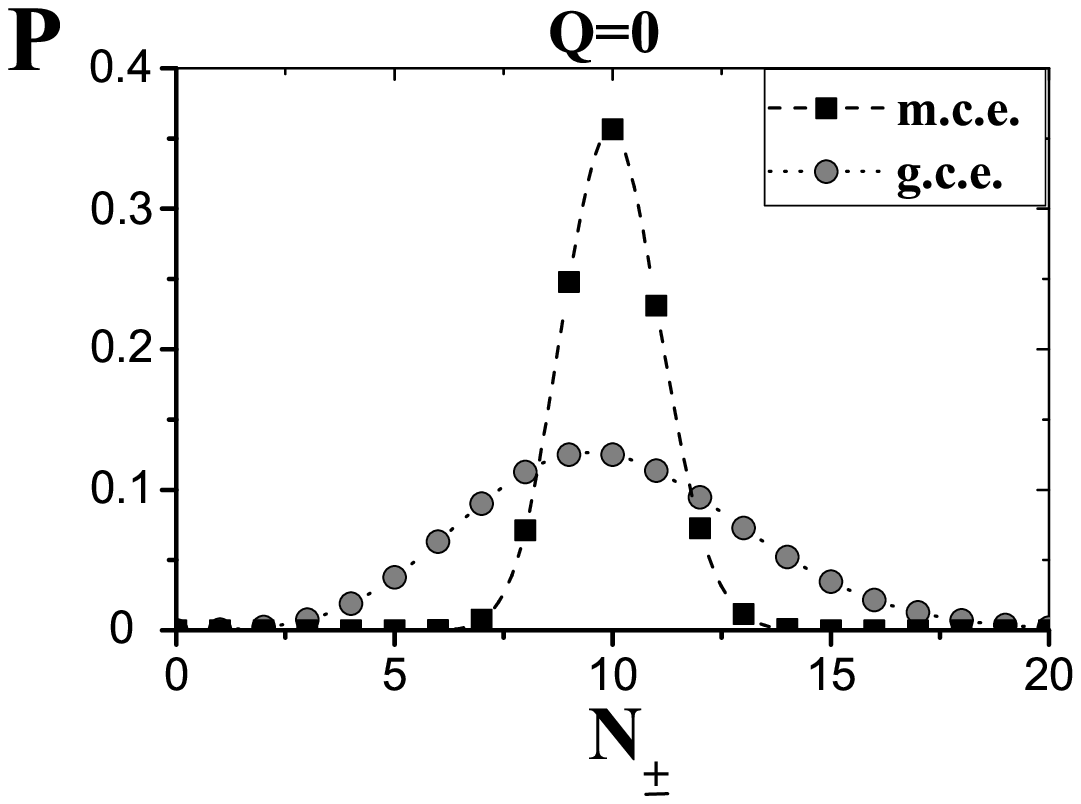,height=12.5cm,width=15cm}
 \vspace{-1.1cm}
 \caption{
 The same as in Fig.~\ref{fig7}, but
 for $~\overline{N}_{\pm}=10~$. \label{fig8}}
\end{figure}
In Figs.~\ref{fig7} and \ref{fig8} we present the particle number
distributions $ P_{m.c.e.}(E,V,N_{\pm},Q=0)$ (\ref{PmceQ}) for
$\overline{N}_{\pm}=0.5$ and $\overline{N}_{\pm}=10$, respectively, and
compare them with the GCE Poisson distributions which remain
 the same as for the case of neutral particles.
Similar to Eqs.~(\ref{int1}-\ref{gaussN}) we find the Gauss approximation
for the MCE particle number distribution in the thermodynamic limit:
\begin{align}
P_{m.c.e.}(E,V,N_{\pm},Q=0) \;\propto\; \exp\left[~-~\frac{4~(N_{\pm} -
\overline{N}_{\pm})^2}
    {\overline{N}_{\pm}}\right]~,
\end{align}
from which it evidently follows $~\langle
N_{\pm}^{2}\rangle_{m.c.e.}-\langle
N_{\pm}\rangle_{m.c.e.}^{2}=\overline{N}_{\pm}/8$ and
$\omega_{m.c.e.}^{\pm} =1/8$.
%
%

\section{Particle number fluctuations for bosons and fermions}

 To study the effects of quantum statistics in the MCE we use the
 technique proposed in Ref.~\cite{steph}.  This
 method allows us to calculate the particle number
fluctuations in the systems with the exact conservation laws imposed in
the thermodynamic limit $V\rightarrow \infty$. We reproduce the
 results of the previous sections for massless particles
 with Boltzmann statistics and study the MCE particle number fluctuations
 in  a general case of the
system of massive charged particles with Bose and/or Fermi statistics
taken into account.

Let us start with a system of neutral bosons or fermions\footnote{There
are examples of neutral bosons, like photon, $\pi^{0}$, $\rho^{0}$, etc.,
which are identical to their antiparticles. For fermions such a
consideration has only illustrative purposes. One always needs to
introduce some kind of charge to distinguish fermions from their
antiparticles.}, then we extend our formulation to a system of charged
particles. The system of neutral non-interacting identical Bose or Fermi
particles can be characterized by the occupation numbers $n_{p}$ of single
quantum states labeled by momenta $p$. The occupation numbers run over
$n_{p}=0,1$ for the fermions and $n_{p}=0,1,2,\dots $ for the bosons. The
GCE average values and fluctuations of $n_{p}$ equal to \cite{landau}:
\eq{
\label{np-aver} \langle n_p \rangle_{g.c.e.} ~&=~
\frac{1}{\exp\left(\epsilon_{p}/T\right)~-~ \gamma}~,\\
\label{np-fluc} \langle (\Delta n_p)^2 \rangle_{g.c.e.}~&\equiv~\langle
n_{p}^{2}\rangle_{g.c.e.}~-~ \langle n_{p}\rangle_{g.c.e.}^{2}~=~ \langle
n_p \rangle_{g.c.e.} \left( 1 ~+~\gamma \langle n_p \rangle_{g.c.e.}
\right)~ \equiv ~v_p^2~, }
 where $\gamma=+1$ and $\gamma=-1$ for Bose and
Fermi statistics, respectively, $\epsilon_{p}\equiv \sqrt{p^2+m^2}$ and
$m$ is the particle mass. Note that $\gamma=0$ in
Eqs.~(\ref{np-aver}--\ref{np-fluc}) corresponds to the Boltzmann
approximation which is valid if $\langle n_{p}\rangle_{g.c.e.}\ll 1$ for
all $p$.

Expressions (\ref{np-aver}-\ref{np-fluc}) are microscopic in the sense
that they describe the average values and fluctuations of single modes
with momentum $p$. However, the fluctuations of macroscopic quantities of
the system can be determined
  through the fluctuations of these single modes. To
  be more precise, we will demonstrate that the particle number
  fluctuations can be written
  in terms of the microscopic correlator $\langle \Delta n_p \Delta
  n_k \rangle_{g.c.e.}$, where $\Delta n_{p}\equiv n_{p}-\langle
  n_{p}\rangle_{g.c.e.}$.
  This correlator can be presented as:
  \eq{
  \langle \Delta n_p \Delta n_k \rangle_{g.c.e.}~ =~
  v_p^{2}~
  \delta_{pk}~,  \label{correlator1}
  }
  due to the fact that the GCE fluctuations of the occupation numbers for different
  quantum states $p\neq k$ are statistically independent.
 The variance $\langle(\Delta N)^2\rangle_{g.c.e.} \equiv
 \langle N^{2}\rangle_{g.c.e.} - \langle N \rangle^{2}_{g.c.e.}$
 of the total number of particles,
  $N \equiv \sum_p n_p$, equals to:
 \eq{\langle (\Delta N)^2\rangle_{g.c.e}~
 =~ \sum_{p,k}~ \langle n_p n_k \rangle_{g.c.e.} -
  \langle n_p \rangle_{g.c.e.} \langle n_k \rangle_{g.c.e.}~
  =~ \sum_{p,k} \langle \Delta n_{p} \Delta n_k
 \rangle_{g.c.e.}~ = ~\sum_p v_p^{ 2}~.
}
 We have assumed above that the quantum $p$-levels
 are non-degenerate. In fact each this level should be
 further specified by the the projection of a particle
 spin $j$. Thus, each $p$-level splits into $g=2j+1$
 sub-levels. It will be assumed that the $p$-summation includes all
 sub-levels too.
 This does not change the above formulation
 because of statistical independence of these quantum sub-levels.
 The degeneracy factor $g$ enters explicitly when one substitutes, in the
 thermodynamic limit,
the summation over discrete levels by the integration:
\eq{\sum_{p}~...~\simeq~\frac{gV}{2\pi^{2}}\int_{0}^{\infty}p^{2}dp~...~.}
The scaled variance $\omega_{g.c.e.}$ in the thermodynamic limit
$V\rightarrow\infty$ reads:
 \eq{\omega_{g.c.e.} ~ \equiv~
 \frac{\langle \left(\Delta N^2\right)\rangle_{g.c.e.}}
 {\langle N \rangle_{g.c.e.}}~
=~ \frac{\sum_{p,k} \langle
 \Delta n_{p} \Delta n_{k} \rangle_{g.c.e.}}
 {\sum_p \langle n_p\rangle_{g.c.e.}}~ = ~
 \frac{\sum_{p}v_p^{ 2}}{\sum_{p}\langle n_p\rangle_{g.c.e.}}~\simeq~
\frac{\int_{0}^{\infty}p^{2}dp~v_p^{ 2}}{\int_{0}^{\infty}p^{2}dp~\langle
n_p\rangle_{g.c.e. } }~. \label{omega-gce}
 }
%

The formula for the microscopic correlator will be modified if we impose
the exact conservation laws in our equilibrated system. We introduce the
equilibrium probability distribution $W(n_p)$ of the different sets
$\{n_{p}\}$ of the occupation numbers.
In the GCE each $n_{p}$ fluctuates independently according approximately
 to the Gauss distribution law for $\Delta n_{p}$ with mean square deviation $v_p^{ 2}$:
 \eq{
   W(n_p) ~\propto ~ \prod_{p}
   \exp{\left[-~ \frac{\left(\Delta n_p\right)^2}{2v_p^{ 2}}
    \right]}~.
\label{gauss} }
To justify Eq.~(\ref{gauss}) one can consider (see Ref.~\cite{steph})
  the sum of $n_p$ in small momentum volume $(\Delta  p)^3$
  with the center at $p$. At fixed $(\Delta p)^3$
  and $V\rightarrow \infty$ the average number of particles
  inside $(\Delta p)^3$ becomes large.
  Each particle configuration inside $(\Delta p)^3$
   consists of $(\Delta p)^3 \cdot gV/(2\pi)^{3}>>1$ statistically
  independent terms, each
with average value $\langle n_{p}\rangle_{g.c.e.}$ (\ref{np-aver}) and
variance $v^{2}_{p}$ (\ref{np-fluc}). From the central limit
  theorem it then follows that the probability distribution for the fluctuations
  inside $(\Delta p)^3$ should be Gaussian.
In fact, we always convolve $n_p$ with some smooth
  function  of $p$, so instead of writing the Gaussian
  distribution for the sum of $n_p$ in $(\Delta p)^3$
we can use it directly for $n_{p}$.

Now we want to impose the exact conservation laws. The conserved quantity
$A$ (the energy and/or conserved charge) can be written in the form
$A\equiv\sum_{p}a(p)\,n_{p}$. An exact conservation law means the
restriction on the sets $\{n_{p}\}$ of the occupation numbers: only those
sets which satisfy the condition $\Delta A=\sum_{p} a(p)\Delta n_p=0$ can
be realized. Let us consider an exact energy conservation. Then
$\;A\rightarrow E\;$ (i.e. $\;a(p)\rightarrow\epsilon_p\;$) and the
distribution (\ref{gauss}) will be modified because of the exact energy
conservation as:
\begin{align}\label{gauss-Q}
   W(n_p) ~\propto ~\prod_{p}
   \exp\left[-~ \frac{\left(\Delta n_p\right)^2}{2v_p^{ 2}}
      \right]~
    \delta\left(\sum_{p}\epsilon_p \Delta n_p^{}
    \right)
 \propto ~  \int_{-\infty}^{\infty} d \lambda~\prod_{p}
  \exp\left[-~ \frac{\left(\Delta n_p\right)^2}{2v_p^{ 2}}
+ i \lambda~\epsilon_p \Delta n_p \right]~,
\end{align}
where $\;\delta\left(\epsilon_p \Delta n_p\right)\;$ is the Dirac
delta-function.
 It is convenient to generalize distribution (\ref{gauss-Q})
 using further the integration along imaginary axis in
 $\lambda$-space.
After completing squares one gets:
   \eq{ W(n_p, \lambda)~ \propto ~  \prod_{p}
    \exp\left[~-
    ~\frac{\left(\Delta n_p - \lambda v_p^{ 2}\epsilon_p\right)^2}
    {2v_p^{ 2}}~+~
    \frac{\lambda^2}{2} v_p^{ 2}\epsilon_p^2\right]~,\label{W-lambda}
    }
and the average values (i.e. the MCE averages) are now calculated as:
\eq{\langle ... \rangle_{m.c.e.}~=~\frac{\int_{-i\infty}^{i\infty}d\lambda
\int_{-\infty}^{\infty}\prod_{p} dn_p^{}~... ~W(n_p,
\lambda)}{\int_{-i\infty}^{i\infty}d\lambda
\int_{-\infty}^{\infty}\prod_{p} dn_p~W(n_p, \lambda)}~. \label{average} }
    Using Eq.~(\ref{average}) one easily deduces
    \eq{
    \langle(\Delta n_p -
    v_p^{ 2}\lambda \epsilon_p)(\Delta n_k - v_k^{ 2}\lambda
    \epsilon_k)   \rangle_{m.c.e.} = \delta_{pk}~ v_p^{ 2}~,
    ~~~
%
   \langle \lambda^2 \rangle_{m.c.e.} = - \left( \sum_{p}
   v_p^{ 2} \epsilon_p^2 \right)^{-1},~~~
%
  \langle (\Delta n_p^{} - v_p^{ 2}\lambda \epsilon_p) \lambda
  \rangle_{m.c.e.}= 0~.\nonumber
  }
 Therefore, one finds the MCE average for the microscopic correlator
  \begin{align}\label{mce-corr}
   \langle \Delta n_p \Delta n_k \rangle_{m.c.e.} ~& =~
   \delta_{pk}~ - ~ v_p^{ 2} \epsilon_p~
   v_k^{ 2} \epsilon_k~ \langle \lambda^2 \rangle ~ + ~
   \langle \Delta n_p \lambda \rangle~ v_k^{ 2}
   \epsilon_k ~ +~ \langle \Delta n_k \lambda \rangle~ v_p^{ 2}
   \epsilon_p
   \\
   &=~ \delta_{pk}~  +~ v_p^{ 2} \epsilon_p~
   v_k^{ 2} \epsilon_k ~\langle \lambda^2 \rangle~
   = ~\delta_{pk}~  v_p^{ 2}~ -~
   \frac{v^{ 2}_p \epsilon_p~v^{ 2}_k  \epsilon_k}
   {\sum_{p} v^{ 2}_p \epsilon_p^2}~. \nonumber
  \end{align}
   By means of Eq.~(\ref{mce-corr}) one obtains:
   \eq{ \label{omega-mce} \omega_{m.c.e.}~&\equiv~\frac{
    \langle(\Delta N^2) \rangle_{m.c.e.} }
    {\langle N \rangle_{m.c.e.}} ~=~ \frac{\sum_{p,k}
    \langle \Delta n_p \Delta n_k \rangle_{m.c.e.}}{\sum_{p}\langle n_{p}\rangle_{m.c.e.}}~
    \simeq~\frac{
    \sum_p v_{p}^{ 2}}{\sum_{p}\langle n_{p}\rangle_{g.c.e.}}~-~
    \frac{\left(\sum_p v_{p}^{ 2}\epsilon_p\right)^{2}}
    {\sum_{p}\langle n_{p}\rangle_{g.c.e.}~\sum_{p} v_{p}^{
    2}\epsilon_p^2} \\
    ~&\simeq ~ \frac{
    \int_{0}^{\infty}p^{2}dp~ v_{p}^{2}}{\int_{0}^{\infty}p^{2}dp~
    \langle n_{p}\rangle_{g.c.e.}}~-~
    \frac{\left(\int_{0}^{\infty}p^{2}dp~ v_{p}^{2}\epsilon_p\right)^{2}}
    {\int_{0}^{\infty}p^{2}dp~\langle n_{p}\rangle_{g.c.e.}~\int_{0}^{\infty}p^{2}dp~
     v_{p}^{2}\epsilon_p^2
    } ~.\nonumber
   }

Comparing Eq.~(\ref{mce-corr}) and Eq.~(\ref{correlator1}) one notices two
changes of the microscopic correlator due to the exact energy
conservation. First, the MCE fluctuations of each mode is reduced, i.e.
the value of $\langle \left(\Delta n_{p}\right)^{2}\rangle_{m.c.e.}$
calculated from Eq.~(\ref{mce-corr}) at $p=k$ is smaller than that of
$\langle \left(\Delta n_{p}\right)^{2}\rangle_{g.c.e.}=v_{p}^{2}$, given
by Eq.~(\ref{np-fluc}) in the GCE. Second, in the MCE the anticorrelations
appear between different modes $p\neq k$ (they are absent in the GCE).
Both these changes result in the MCE suppression of the scaled variances
(\ref{omega-mce}) in a comparison with those in the GCE (\ref{omega-gce}).
In fact, the first term in the r.h.s. of Eq.~(\ref{omega-mce}) equals to
the GCE scaled variance $\omega_{g.c.e.}$ (\ref{omega-gce}) and the second
negative term corresponds to the MCE suppression effects. Note also that
according to Eq.~(\ref{omega-mce}) the MCE fluctuations in the
thermodynamic limit $V\rightarrow\infty$ can be presented in terms of the
GCE quantities. The exact energy conservation should also lead to the
differences between $\langle n_{p}\rangle_{m.c.e.}$ and $\langle
n_{p}\rangle_{g.c.e.}$. However, in the thermodynamic limit it follows:
$\sum_{p}\langle n_{p}\rangle_{m.c.e.}\simeq \sum_{p}\langle
n_{p}\rangle_{g.c.e.}$. It can be also proven by the straightforward
calculations that differences between $\langle n_{p}\rangle_{m.c.e.}$ and
$\langle n_{p}\rangle_{g.c.e.}$ lead to the corrections of the order of
$\overline{N}$ to both $\langle N^{2}\rangle_{m.c.e.}$ and $\langle
N\rangle_{m.c.e.}^{2}$, but these corrections are equal to each other and
they are cancelled out in the calculation of $\langle\left(\Delta
N\right)^{2}\rangle_{m.c.e.}$.

\begin{figure}[ht]
\epsfig{file=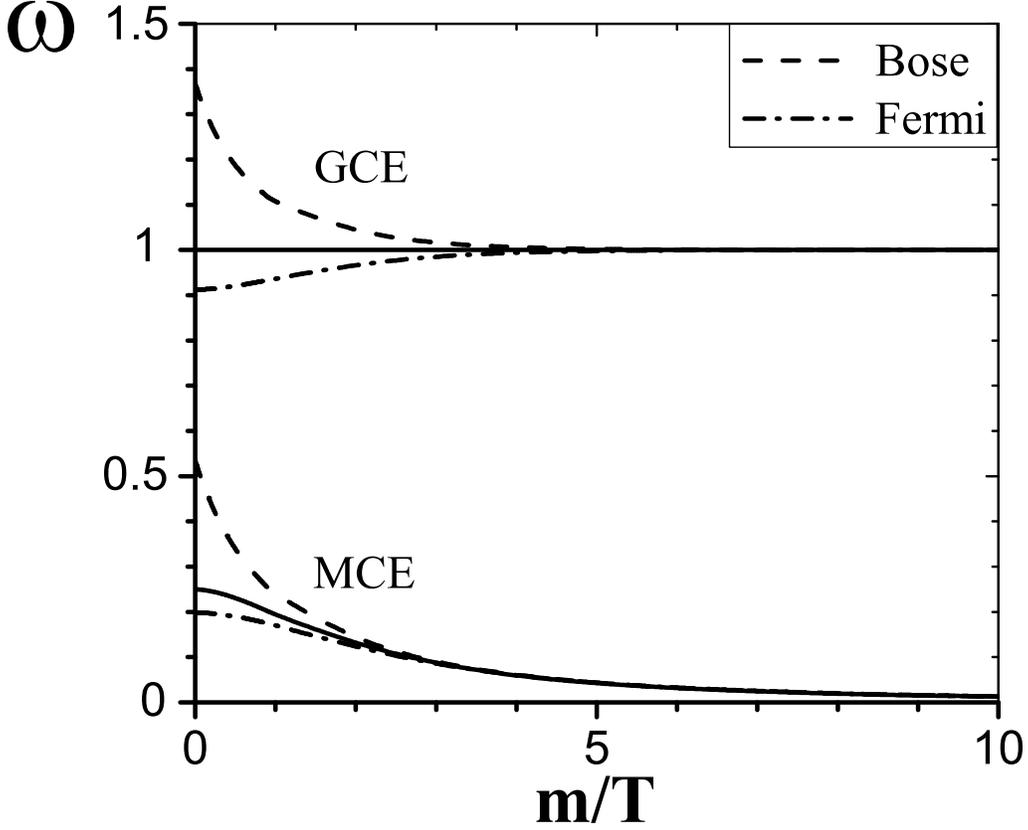,height=12cm,width=14cm}
  \caption{Three upper lines present the GCE scaled variances
  $\omega_{g.c.e.}$ (\ref{omega-gce}) at different values of $m/T$,
  whereas three lower lines
  correspond to the MCE scaled variances $\omega_{m.c.e.}$ (\ref{omega-mce}).
 The dashed lines correspond to the Bose statistics,
the dashed-dotted lines to the Fermi statistics and the solid lines to the
Boltzmann approximation.
 }\label{fig9}
\end{figure}

The GCE (\ref{omega-gce}) and MCE (\ref{omega-mce}) scaled variances for
different statistics are shown as functions of $m/T$ in Fig.~\ref{fig9}.
In the Boltzmann approximation ($\gamma=0$) from Eq.~(\ref{omega-gce}) one
finds:
\eq {\omega_{g.c.e.}^{Boltz}~=~1~, \label{omega-Boltz}
}
which coincides with Eq.~(\ref{omegagce}) used in Section III for $m=0$.
The Eq.~(\ref{omega-Boltz}) remains valid for all values of $m/T$. This is
due to the fact that it follows from the Poisson particle number
distribution $P_{g.c.e.}^{Boltz}(\overline{N},N)$ in the GCE, which is
given by Eq.~(\ref{PNgce}) at all values of $m/T$ (only the average value
of particle number $\overline{N}$ decreases with increasing of $m/T$).
From Eq.~(\ref{omega-gce}) it follows that the effects of quantum
statistics lead to the Bose enhancement, $\omega^{Bose}_{g.c.e.}>1$ at
$\gamma=1$, and the Fermi suppression, $\omega^{Fermi}_{g.c.e.}<1$ at
$\gamma=-1$, of the particle number fluctuations. The strongest quantum
statistic effects correspond to the $m/T\rightarrow 0$ limit:
 \eq{
\omega_{g.c.e.}^{Bose}(m=0)~&=~\frac{\pi^{2}}{6\,\zeta(3)} ~\simeq
~1.368~,\label{omegaB}\\
\omega_{g.c.e.}^{Fermi}(m=0)~&=~\frac{\pi^{2}}{9\,\zeta(3)}~ \simeq~
0.912~. \label{omegaF} }

For the particle number fluctuations, as seen from Fig.~\ref{fig9}, the
Bose enhancement $\omega^{Bose}/\omega^{Boltz}>1$ and the Fermi
suppression $\omega^{Fermi}/\omega^{Boltz}<1$ factors decrease
monotonously with increasing of $m/T$, in both the GCE and the MCE. These
effects of quantum statistics in both ensembles become negligible at
$m/T\gg 1$, as in this limit one finds
%
$\langle n_{p}\rangle_{g.c.e.} \simeq \exp(-\epsilon_{p}/T)\ll 1$, so that
$v_{p}^{2}\simeq \langle n_{p}\rangle_{g.c.e.}$ for both the Bose and
Fermi statistics.

For the Boltzmann approximation all momentum integrals in
Eq.~(\ref{omega-mce}) for $\omega_{m.c.e.}$ can be calculated
analytically:
\begin{align}
& \int_0^{\infty}p^{2}dp~ \exp\left(-\frac{\epsilon_{p}}{T}\right)
  \;=\; T\,m^2K_2\left(\frac{m}{T}\right)~,~~~
\int_0^{\infty} p^{2} dp~
\epsilon_{p}~\exp\left(-\frac{\epsilon_{p}}{T}\right)\; =\;
\frac{m^4}{8}\left[K_4\left(\frac{m}{T}\right)~-~
K_0\left(\frac{m}{T}\right)\right]~,\nonumber
  \\
& \int_0^{\infty} p^{2} dp~
\epsilon_{p}^{2}~\exp\left(-\frac{\epsilon_{p}}{T}\right)\; =\;
\frac{m^5}{16}\left[K_5\left(\frac{m}{T}\right)~
    + ~K_3\left(\frac{m}{T}\right)~ - ~2\,K_1\left(\frac{m}{T}\right)
    \;\right]\;. \nonumber
\end{align}
Making use of the asymptotic behavior of the modified Hankel function
$\;K_n(x)\;$ at $x\rightarrow 0$ ($K_0(x) \simeq-\ln x\;$ and
$K_n(x)\simeq
 \frac{1}{2}\;\Gamma(n)\left(\frac{x}{2}\right)^{-n}\;$ for $n\geq 1$)
one gets in the $m/T\rightarrow 0$ limit:
\begin{align}\label{omega-mce-Boltz}
\omega_{m.c.e.}^{Boltz}(m=0)~=~\frac{1}{4}~,
\end{align}
i.e. for classical massless particles the MCE scale variance is quarter as
large as the corresponding scaled variance in the GCE. This result
coincides, of course, with that of Eq.~(\ref{omegas}) obtained in Section
III from the MCE partition function of massless particles with Boltzmann
statistics. For the case of Bose and Fermi statistics we obtain:
\begin{align}
\omega_{m.c.e.}^{Bose}(m=0)~&=~ \frac{\pi^2}{6\,\xi(3)}~ -~
\frac{135\,\xi(3)}{2 \pi^4} ~\simeq ~0.535
\;, \label{omega-mce-B}\\
\omega_{m.c.e.}^{Fermi}(m=0)~&= ~\frac{\pi^2}{9\,\xi(3)}~ - ~\frac{405\,
\xi(3)}{7 \pi^4} ~\simeq~ 0.198 \;. \label{omega-mce-F}
\end{align}
%
The Bose enhancement and Fermi suppression of the fluctuations exist in
both the GCE and the MCE. However,
the effects of quantum statistics for the particle number fluctuations are
stronger in the MCE than those in the GCE . As it follows from
Eqs.~(\ref{omega-mce-Boltz}--\ref{omega-mce-F})
 the Bose enhancement of the scaled
variance in the MCE at $m/T\rightarrow 0$ is approximately equal to 2.142,
and the Fermi suppression 0.793. These numbers can be compared with 1.368
and 0.912, respectively, found from Eqs.~(\ref{omega-Boltz}--\ref{omegaF})
in the GCE.

As seen from Fig.~\ref{fig9} the MCE suppression of the particle number
fluctuations for massive particles is stronger than that for massless
ones, and all MCE scaled variances $\omega_{m.c.e.}$ decrease monotonously
with increasing of $m/T$. From the asymptotic expansion at $x\gg 1$
\cite{I}
\eq{ K_n(x)~\simeq ~\sqrt{\pi/2x}\,\exp(-x)~\sum_{k\geq 0}
\frac{1}{(2x)^{k}}~\frac{(4n^2-1)...[4n^{2}-(2k-1)^{2}]}{k!\,2^{2k}}
\label{Kn}
}
one finds the behavior of the scaled variance
$\omega_{m.c.e.}^{Boltz}\simeq \frac{3}{2}(m/T)^{-2}\ll 1$ at $m/T\gg 1$
for the Boltzmann approximation (to obtain this result one needs to keep
terms up to $k=3$ in Eq.~(\ref{Kn})). The MCE scaled variances for the
Bose and Fermi statistics have the same behavior, as the effects of
quantum statistics
are negligible at $m/T\gg 1$, so that $\omega_{m.c.e.}^{Bose}\simeq
\omega_{m.c.e.}^{Fermi} \simeq \omega_{m.c.e.}^{Boltz}$.


\vspace{0.3cm} For the system of charged particles and antiparticles in
the GCE, similar to Eqs.~(\ref{np-aver}-\ref{np-fluc}), one has:
\eq{ \label{np-aver-ch}
\langle n_p^{\pm} \rangle_{g.c.e.} ~&=~
\frac{1}{\exp\left[\left(\epsilon_{p}~\mp~\mu\right)/T\right]~-~ \gamma}~, \\
\label{np-fluc-ch} \langle (\Delta n_p^{\pm})^2 \rangle_{g.c.e.}~&
\equiv~\langle \left(n_{p}^{\pm}\right)^{2}\rangle_{g.c.e.}~-~ \langle
n_{p}^{\pm}\rangle_{g.c.e.}^{2}~ =~ \langle n_p^{\pm} \rangle_{g.c.e.}
\left( 1~ +~\gamma \langle n_p^{\pm} \rangle_{g.c.e.} \right)~ \equiv
~v_p^{\pm 2}~, }
 where $\mu$ is the chemical potential connected with the
conserved charge $Q$. The microscopic correlator (\ref{correlator1}) can
be then generalized as:
 \eq{
  \langle \Delta n_p^{\alpha} \Delta n_k^{\beta} \rangle_{g.c.e.}~ =~
  v_p^{\alpha 2}~  \delta_{pk}~\delta_{\alpha \beta}~,  \label{corr-gce-ch}
  }
  where $\alpha, \beta$ are + and(or) $-$.
The average values of the energy $E=\sum_{p,\alpha}
\epsilon_{p}n_{p}^{\alpha}$ and charge\footnote{In what follows we assume
$q^{+}=1$ and $q^{-}=-1$, so that $Q=N_{+}-N_{-}$. However, other values
with $q^{+}=-q^{-}$ can be also used.} $Q=\sum_{p,\alpha}
q^{\alpha}n_{p}^{\alpha}$ are regulated in the GCE by the temperature $T$
and the chemical potential $\mu$, respectively. Similar to
Eq.~(\ref{gauss-Q}) the MCE distribution for the occupation numbers
$n_{p}^{\alpha}$ can be presented as:
\begin{align}\label{gauss-QE}
   W(n_p) ~&\propto ~\prod_{p,\alpha}
   \exp\left[-~ \frac{\left(\Delta n_p^{\alpha}\right)^2}{2v_p^{\alpha 2}}
      \right]~
    \delta\left(\sum_{p,\alpha}\epsilon_p \Delta n_p^{\alpha}
    \right)~
    \delta\left(\sum_{p,\alpha}q^{\alpha} \Delta n_p^{\alpha}
    \right)\\
 & \propto ~  \int_{-\infty}^{\infty} d \lambda_{E}~
 \int_{-\infty}^{\infty} d \lambda_{Q}~\prod_{p,\alpha}
~ \exp\left[~-~ \frac{\left(\Delta n_p^{\alpha}\right)^2} {2v_p^{\alpha\,
2}}~ +~ i\, \lambda_{E}\,\epsilon_p \Delta n_p^{\alpha} ~+~i\,
\lambda_{Q}\,q^{\alpha} \Delta n_p^{\alpha} \right]~,\nonumber
\end{align}
where $\delta\left(\sum_{p,\alpha}\epsilon_p \Delta n_p^{\alpha}
    \right)$ and $\delta\left(\sum_{p,\alpha}q^{\alpha} \Delta n_p^{\alpha}
    \right)$ correspond to the exact energy and charge conservations,
    respectively, in the MCE.
After the straightforward calculations, similar to
Eqs.~(\ref{W-lambda}--\ref{mce-corr}), one gets an expression for the
microscopic correlator in the MCE for the charged particles with the exact
charge conservation law imposed:
   \eq{\label{mce-corr-ch}
   \langle \Delta n_p^{\alpha} \Delta
   n_k^{\beta} \rangle_{m.c.e.}~ =~ v^{\alpha 2}_p \delta_{pk}\delta_{\alpha
   \beta}
   ~-~
   \frac{v^{\alpha 2}_p v^{\beta 2}_k}{|A|}
   \left[
   q^{\alpha}q^{\beta}\sum_{p,\alpha}v^{\alpha 2}_p \epsilon_p^2 ~ + ~
   \epsilon_p \epsilon_k \sum_{p,\alpha} v^{\alpha 2}_p q^{\alpha 2}
    ~-~\left(q^{\alpha}\epsilon_k+q^{\beta}\epsilon_p\right)
   \sum_{p,\alpha}v^{\alpha 2}_p\epsilon_p
   q^{\alpha}\right]~,
   }
   where
   \eq{
   |A|~ \equiv ~ \left(\sum_{p,\alpha}v^{\alpha 2}_p \epsilon_p^2\right)
   \cdot \left(\sum_{p,\alpha}
   v^{\alpha 2}_p  q^{\alpha 2}\right)~ - ~\left(\sum_{p,\alpha}v^{\alpha 2}_p\epsilon_p
   q^{\alpha}\right)^2~.
   }
Therefore, one finds:
   \eq{ \label{omega-mceQ}
   \omega_{m.c.e.}^{\alpha}~&\equiv~\frac{
    \langle(\Delta N_{\alpha}^{2}) \rangle_{m.c.e.} }
    {\langle N_{\alpha} \rangle_{m.c.e.}} ~=~ \frac{\sum_{p,k}
    \langle \Delta n_p^{\alpha} \Delta n_k^{\alpha} \rangle_{m.c.e.}}
    {\sum_{p}\langle n_{p}^{\alpha}\rangle_{m.c.e.}}
    \\
 &\simeq ~\frac{
    \sum_p v_{p}^{\alpha\, 2}}{\sum_{p}\langle n_{p}^{\alpha}\rangle_{g.c.e.}}
    ~-~\frac{\left(\sum_p v_{p}^{\alpha\, 2}\epsilon_p\right)^{2}}
    {\sum_{p}\langle n_{p}^{\alpha}\rangle_{g.c.e.}~
    \sum_{p,\alpha}v^{\alpha 2}_p \epsilon_p^2}
    ~-~\frac{
    \left(\sum_{p}v^{\alpha 2}_p  q^{\alpha}\right)^2}
    {\sum_{p}\langle n_{p}^{\alpha}\rangle_{g.c.e.}
    ~\sum_{p,\alpha}v^{\alpha 2}_p  q^{\alpha 2}}
    \;.\nonumber
   }
The first term in the r.h.s. of Eq.~(\ref{omega-mceQ}) corresponds to the
GCE scaled variance $\omega_{g.c.e.}^{\alpha}$ for the positive
($\alpha=1$) or negative ($\alpha=-1$) particles. Note that in the
Boltzmann approximation $\omega_{g.c.e.}^{\alpha ~Boltz}$ coincides with
the scaled variance for the neutral particles $\omega_{g.c.e.}^{Boltz}$
(\ref{omega-Boltz}) and, therefore, equals 1. The second and third terms
in the r.h.s. of Eq.~(\ref{omega-mceQ}) correspond to the MCE suppression
of the fluctuations due to the exact conservations of energy (compare to
Eq.~(\ref{omega-mce})) and charge, respectively.

In the case of zero net charge $Q=0$ (this means $\mu=0$ and, therefore,
$\langle n_{p}^{+}\rangle_{g.c.e.}=\langle n_{p}^{-}\rangle_{g.c.e.}$) one
finds for charged particles:
 \eq{
\omega_{m.c.e.}^{\pm}(Q=0) \;=\;
 \frac{\sum_p\upsilon_p^{\pm\;2}}{\sum_p \langle n_p^{\pm}\rangle_{g.c.e.}}
 \;-\; \frac{1}{2}\;
 \frac{\sum_p\upsilon_p^{\pm\;2}}
      {\sum_p \langle n_p^{\pm}\rangle_{g.c.e.}}
 \;-\; \frac{1}{2}\;
 \frac{\left(\sum_p\upsilon_p^{\pm\;2}\epsilon_p\right)^2}
      {\sum_p \langle n_p^{\pm}\rangle_{g.c.e.}
       ~\sum_p\upsilon_p^{\pm\;2}\epsilon_p^2}
 \;=\;\frac{1}{2}\;\omega_{m.c.e.}\;, \label{omega-mce-0}
 }
where $\omega_{m.c.e.}$ in Eq.~(\ref{omega-mce-0}) corresponds to the MCE
scaled variance (\ref{omega-mce}) for the neutral particles with the
parameters $m$ and $g$ equal to those of the charged particles. Thus, we
came to the conclusion that in the MCE at $Q=0$ an exact charge
conservation leads to the scaled variances of (negative) positive
particles which are by a factor of 1/2 smaller than the corresponding MCE
scaled variances for neutral particles. This result is an agreement with
the CE suppression of the particle number fluctuations found in
Ref.~\cite{ce-fluc}: the scaled variance in the CE at $Q=0$ in the
thermodynamic limit equals to
$\omega^{\pm}_{c.e.}=\frac{1}{2}\,\omega^{\pm}_{g.c.e.}$
(note that at $\mu=0$ the GCE scaled variances for charged and neutral
particles are equal to each other). In the case of Boltzmann statistics
one finds for massless particles
$ \omega_{m.c.e.}^{\pm Boltz}(m=0) \;=\;
    1/8 \;,
$
%
which coincides, of course, with the result of Eq.~(\ref{omegamceQas})
obtained in Sec.~IV.

\section{Summary}
   We have studied the particle number fluctuations
   in the MCE. First, in Sec.~II we have
   obtained the partition function for the system of
   non-interacting massless neutral
   particles with Boltzmann statistics.
   This allows us to calculate the particle number fluctuations
   for this system in Sec.~III.
   These MCE fluctuations are suppressed in comparison to those in the GCE.
   In the thermodynamic limit
   the MCE scaled variance of the multiplicity
   distribution equals  1/4 and this is a quarter the size of  that
   in the GCE.
   As a second step we consider in Sec.~IV the system
   of the charged massless particles with zero net charge and
   calculate its partition function. This leads to the scaled
   variance equal to 1/8 in the thermodynamic limit.
   Thus, an exact charge conservation
   makes an additional suppression of the scaled variance by a factor of $1/2$.
   In Sec.~V we use an approach proposed in
   Ref.~\cite{steph}. This method does not work accurately for finite
   ({\it small}) systems, but it does allow us to calculate the correct
   values of the scaled variances
   in the thermodynamic limit $V\rightarrow\infty$ for a much broader
   class of statistical systems --  massive bosons and fermions --
    when both the energy and charge exact conservations
   are taken into account.
    The effects of quantum statistics and non-zero particle mass
    lead to the significant changes of the MCE particle number
    fluctuations, and they are studied in details  in Sec.~V.
We have also reproduced the limiting $V\rightarrow\infty$ behavior of the
MCE scaled variances for massless particles with Boltzmann statistics
obtained in Secs.~III--IV from the exact analytical expressions for the
partition functions.

\begin{acknowledgments}
We are grateful to F.~Becattini, A.I.~Bugrij, M.~Ga\'zdzicki, W.~Greiner,
A.~Ker\"anen, I.N.~Mishustin, St.~Mr\'owczy\'nski, L.M.~Satarov,
Y.M.~Sinyukov, M.~Stephanov and H.~St\"ocker for useful discussions. We
thank A.~Koetsier for help in preparing the manuscript. The work was
   partially supported by US Civilian Research and Development
   Foundation (CRDF) Cooperative Grants Program, Project Agreement
   UKP1-2613-KV-04.
\end{acknowledgments}

\appendix

\section{}

The microcanonical partition function $\;W_N(E,V)\;$ can be
recursively calculated from $\;W_{N-1}(E,V)\;$:
\begin{align}\label{omegaN}
W_N(E,V)
 &\;=\; \frac{1}{N!}\,\frac{gV}{(2\pi )^3}\int d^3p^{(N)}\;
        \frac{gV}{(2\pi )^3}\int d^3p^{(N-1)}\ldots
       \frac{gV}{(2\pi )^3}\int d^3p^{(1)}\;
       \delta(E-\sum_{k=1}^N|\vec{p}^{\;(k)}|)
  \\
 &\;=\; \frac{1}{N}\,4\pi\,\frac{gV}{(2\pi )^3}\int_0^E dp^{(N)}
        \left(p^{(N)}\right)^2\frac{1}{(N-1)!}\,\frac{gV}{(2\pi )^3}
        \int d^3p^{(N-1)}\ldots \nonumber
  \\
 &\qquad\times \frac{gV}{(2\pi )^3}\int d^3p^{(1)}\;
        \delta(E-p^{(N)}-\sum_{k=1}^{N-1}|\vec{p}^{\;(k)}|) \nonumber
  \\
 &\;=\;
        \frac{1}{N}\frac{gV}{2\pi^2}
        \int_0^E dp\,p^2\;W_{N-1}(E-p,V)~.\nonumber
\end{align}
Now we shall prove by  induction method that for arbitrary
$\;N\geq 1\;$ the microcanonical partition function has the form
of Eq.~(\ref{omegaNa}). It can be checked directly that
Eq.~(\ref{omegaNa}) at $\;N=1\;$ and $\;N=2\;$ coincides with
(\ref{omega1}) and (\ref{omega2}), respectively. Now assume that
it is correct for $\;N-1\;$,
\begin{align}\label{WN-1}
W_{N-1}(E,V)
 \;=\; \frac{2^{N-1}}{(3N-4)!(N-1)!}
 \left(\frac{gV}{2\pi^2}\right)^{N-1} E^{3N-4}\;,
\end{align}
and substitute (\ref{WN-1}) into Eq.~(\ref{omegaN}):
\begin{align}\label{omegaNE}
W_N(E,V)
 &\;=\; \frac{1}{N}\,\frac{gV}{2\pi^2} \int_0^E dp\,p^2
        \frac{2^{N-1}}{(3N-4)!(N-1)!}\left(\frac{gV}
        {2\pi^2}\right)^{N-1} (E-p)^{3N-4}
  \\
 &\;=\; \frac{1}{N}\left(\frac{gV}{2\pi^2}
 \right)^N \frac{2^{N-1}}{(3N-4)!(N-1)!}
        \int_0^E dp\,p^2\; (E-p)^{3N-4}~.\nonumber
\end{align}
The last integral can be easily evaluated after substitution $\;p=
E-q\;$:
\begin{align}\label{int}
 \int_0^E dp\,p^2\; (E-p)^{3N-4} \;=\;
 \int_0^E dq\,(E-q)^2\;q^{3N-4} \;=\; \frac{2E^{3N-1}}
 {(3N-1)(3N-2)(3N-3)}~.
\end{align}
Eqs.\;(\ref{omegaNE}-\ref{int}) result in Eq.~(\ref{omegaNa}).

\section{}

The generalized hypergeometric function (GHF), also known as the Barnes
extended hypergeometric functionis defined by the following
series,
\begin{equation}\label{GHF}
{}_p F_q(a_1,a_2,\dots,a_p;b_1,b_2,\dots,b_q;z)=
\sum_{k=0}^{\infty} \frac{(a_1)_k (a_2)_k \dots (a_p)_k~z^k}
{(b_1)_k (b_2)_k \dots (b_q)_k~k!} ~,
\end{equation}
where $(a)_k$ is the Pochhammer symbol:
\begin{equation}\label{Poch}
(a)_k \equiv \frac{\Gamma(a+k)}{\Gamma(a)} = a (a+1) \dots (a+k-1)~.
\end{equation}


The asymptotic behavior of GHF at $z \rightarrow \infty$ is given
at p=0 by
\begin{eqnarray}
{}_0 F_q(;b_1,b_2,\dots,b_q;z) &\simeq&
\frac{\prod_{j=1}^{q} \Gamma(b_j)}{\sqrt{(q+1) (2 \pi)^q}}
(\sqrt[q+1]{z})^{\frac{q}{2}-\sum_{j=1}^{q} b_j}
e^{(q+1) \sqrt[q+1]{z}} \nonumber \\
& &
\left(1+\frac{c_1}{\sqrt[q+1]{z}}+
\frac{c_2}{(\sqrt[q+1]{z})^2} +
\frac{c_3}{(\sqrt[q+1]{z})^3} + \dots \right), \label{GHFas}
\end{eqnarray}
where
\begin{eqnarray}
c_1 &=& - \frac{1}{2} D_2 + \frac{q (q+2)}{24 (q+1)},\\
c_2 &=& \frac{1}{8} D_2^2 + \frac{1}{6} D_3 -
\frac{q(q+2)+12}{48 (q+1)} D_2 +
\frac{q (q+2) \left[ q (q+2)+24 \right]}{1152 (q+1)^2},\\
c_3 &=& - \frac{1}{48} D_2^3 - \frac{1}{12} D_2 D_3
- \frac{1}{12} D_4 + \frac{q(q+2)+48}{48 (q+1)}
\left( \frac{1}{4} D_2^2 + \frac{1}{3} D_3 \right) \\
& & - \frac{q^2(q+2)^2+768}{2304 (q+1)^2} D_2
+\frac{q (q+2) \left[5 q^2 (q+2)^2 -288 q (q+2) + 10 944 \right]}
{414720 (q+1)^3},\nonumber
\end{eqnarray}
\begin{eqnarray}
D_2 &=& B_2 - \frac{1}{(q+1)} B_1^2 ,\\
D_3 &=& B_3 - \frac{3}{(q+1)} B_1 B_2 + \frac{2}{(q+1)^2} B_1^3 ,\\
D_4 &=& B_4 - \frac{4}{(q+1)} B_1 B_3 + \frac{6}{(q+1)^2} B_1^2 B_2
- \frac{3}{(q+1)^3} B_1^4,
\end{eqnarray}
and
\begin{equation}
B_m = 1 + \sum_{j=1}^{q} \left( b_j \right)^m  \ .
\end{equation}


\begin{thebibliography}{10}

\bibitem{stat-model}
J. Cleymans, H. Satz, Z. Phys. C {\bf 57} (1993) 135; J.
Sollfrank, M. Ga\'zdzicki, U. Heinz, J. Rafelski, Z. Phys. C {\bf
61} (1994) 659; G.D. Yen, M.I. Gorenstein, W. Greiner, S.N.~Yang,
Phys. Rev. C {\bf 56} (1997) 2210; F.~Becattini, M. Ga\'zdzicki, J.
Solfrank, Eur. Phys. J. C {\bf  5} (1998) 143; G.D. Yen, M.I.
Gorenstein, Phys. Rev. C {\bf  59} (1999) 2788; P. Braun-Munzinger,
I. Heppe, J. Stachel, Phys. Lett. B {\bf  465} (1999) 15;
P.~Braun-Munzinger, D. Magestro, K. Redlich, J. Stachel, Phys.
Lett. B {\bf  518} (2001) 41; F.~Becattini, M. Ga\'zdzicki,
A.~Keranen, J.~Manninen, R.~Stock, Phys. Rev. C {\bf 69} (2004)
024905.

\bibitem{PBM}
P. Braun-Munzinger, K. Redlich, J. Stachel, nucl-th/0304013,
Review for Quark Gluon Plasma 3, eds. R. C. Hwa and Xin-Nian Wang,
World Scientific Publishing.

\bibitem{ce}
K. Redlich, L. Turko, Z. Phys. C {\bf  5} (1980) 541; J. Rafelski,
M. Danos, Phys. Lett. B {\bf  97} (1980) 279. J. Cleymans, K
Redlich, E Suhonen, Z. Phys. {\bf C 51} (1991) 137; J. Cleymans,
A. Ker\"anen, M. Marais, E. Suhonen, Phys. Rev. C {\bf  56} (1997)
2747; F.~Becattini, Z. Phys. C {\bf  69} (1996) 485; F. Becattini,
U. Heinz, Z. Phys. C {\bf  76} (1997) 269; J. Cleymans, H.
Oeschler, K. Redlich, Phys. Rev. C {\bf  59} (1999) 1663; Phys.
Lett. B {\bf  485} (2001) 27; M.I. Gorenstein, M. Ga\'zdzicki, W.
Greiner, Phys. Lett. B {\bf  483} (2000) 60; M.I. Gorenstein, A.P.
Kostyuk, H. St\"ocker, W. Greiner, Phys. Lett. B {\bf 509} (2001)
277.

\bibitem{mce}
K. Werner, J. Aichelin, Phys. Rev. C {\bf 52} (1995) 1584;
F.M.~Liu, K.~Werner, J.~Aichelin, Phys. Rev. C {\bf 68} (2003)
024905;
 F.~Liu, K.~Werner, J.~Aichelin, M.~Bleicher,
 H.~St\"ocker, J. Phys. G {\bf
30} (2004) S589; F.~Becattini, L.~Ferroni, Eur. Phys. J. C
{\bf  35} (2004) 243;
hep-ph/0407117.

\bibitem{fluc} Misha A. Stephanov, K Rajagopal, Edward V. Shuryak,
Phys. Rev. Lett. {\bf 81} (1998) 4816; Henning Heiselberg, Phys. Rep. {\bf
351} (2001) 161;
S. Jeon, V. Koch, hep-ph/0304012,
Review for Quark-Gluon Plasma 3, eds. R.C. Hwa and X.-N. Wang,
World Scientific, Publishing;
 M. Ga\'zdzicki, M.I.
Gorenstein, St.~Mr\'owczy\'nski, Phys. Lett. B {\bf  585} (2004)
115;
M.I.~Gorenstein, M. Ga\'zdzicki, O.S. Zozulya, Phys. Lett. B {\bf 585}
(2004) 237.

\bibitem{ce-fluc} V.V. Begun, M. Ga\'zdzicki, M.I. Gorenstein, O.S. Zozulya,
Phys. Rev. C {\bf 70} (2004) 034901; V.V. Begun, M.I. Gorenstein, O.S.
Zozulya, nucl-th/0411003.

\bibitem{steph} Misha A. Stephanov, K Rajagopal, Edward V. Shuryak,
Phys. Rev. D {\bf 60} (1999) 114028.

\bibitem{I}
M. Abramowitz and I.E. Stegun,
 {\it Handbook of Mathematical Functions} (Dover, New York, 1964).

\bibitem{landau}
L.D. Landau, E.M. Lifschitz, {\it Statistical Physics} (Fizmatlit, Moscow,
2001).

\end{thebibliography}
\end{document}